\documentclass[conference]{IEEEtran}
\IEEEoverridecommandlockouts

\usepackage{cite}
\usepackage{amsmath,amssymb,amsfonts} 
\usepackage{algorithmic}
\usepackage{graphicx}
\usepackage{textcomp}
\usepackage{multirow} 
\usepackage{booktabs}
\usepackage{pdflscape}
\usepackage{threeparttable}
\usepackage[T1]{fontenc} 
\usepackage{soul} 
\usepackage{todonotes}
\usepackage{pgfplots}
\usepgfplotslibrary{fillbetween}
\usepackage[caption=false]{subfig} 
\usepackage{comment}
\pgfplotsset{compat=1.17}

\def\BibTeX{{\rm B\kern-.05em{\sc i\kern-.025em b}\kern-.08em
    T\kern-.1667em\lower.7ex\hbox{E}\kern-.125emX}}

\usepackage[final,commandnameprefix=always]{changes}

\newcommand{\stkout}[1]{\ifmmode\text{\sout{\ensuremath{#1}}}\else\sout{#1}\fi}
\setdeletedmarkup{\stkout{#1}}

\usepackage{amsthm}

\theoremstyle{definition}

\makeatletter
\newcommand{\linebreakand}{%
  \end{@IEEEauthorhalign}
  \hfill\mbox{}\par
  \mbox{}\hfill\begin{@IEEEauthorhalign}
}
\makeatother

\usepackage[edges]{forest}
\definecolor{folderbg}{RGB}{124,166,198}
\definecolor{folderborder}{RGB}{110,144,169}
\newlength\Size
\tikzset{%
  folder/.pic={%
    \filldraw [draw=folderborder, top color=folderbg!50, bottom color=folderbg] (-1.05*\Size,0.2\Size+5pt) rectangle ++(.75*\Size,-0.2\Size-5pt);
    \filldraw [draw=folderborder, top color=folderbg!50, bottom color=folderbg] (-1.15*\Size,-\Size) rectangle (1.15*\Size,\Size);
  },
  file/.pic={%
    \filldraw [draw=folderborder, top color=folderbg!5, bottom color=folderbg!10] (-\Size,.4*\Size+5pt) coordinate (a) |- (\Size,-1.2*\Size) coordinate (b) -- ++(0,1.6*\Size) coordinate (c) -- ++(-5pt,5pt) coordinate (d) -- cycle (d) |- (c) ;
  },
}
\forestset{%
  declare autowrapped toks={pic me}{},
  pic dir tree/.style={%
    for tree={%
      folder,
      font=\ttfamily,
      grow'=0,
    },
    before typesetting nodes={%
      for tree={%
        edge label+/.option={pic me},
      },
    },
  },
  pic me set/.code n args=2{%
    \forestset{%
      #1/.style={%
        inner xsep=2\Size,
        pic me={pic {#2}},
      }
    }
  },
  pic me set={directory}{folder},
  pic me set={file}{file},
}

\begin{document}

\title{Dynamic Fee for Reducing Impermanent Loss in Decentralized Exchanges

\vspace{0cm}}

\author{
\IEEEauthorblockN{
  Irina Lebedeva\IEEEauthorrefmark{1},
  Dmitrii Umnov\IEEEauthorrefmark{2},
  Yury Yanovich\IEEEauthorrefmark{1},
  Ignat Melnikov\IEEEauthorrefmark{1},
  George Ovchinnikov\IEEEauthorrefmark{1}\IEEEauthorrefmark{3}
}
\IEEEauthorblockA{\IEEEauthorrefmark{1}%
  Skolkovo Institute of Science and Technology, Moscow, Russia \\
}
\IEEEauthorblockA{\IEEEauthorrefmark{2}%
  Faculty of Computer Science, HSE University, Moscow, Russia \\
}
\IEEEauthorblockA{\IEEEauthorrefmark{3}%
  Kharkevich Institute for Information Transmission Problems, RAS, Moscow, Russia
}
\vspace{-1cm}
}

\maketitle

\begin{abstract}


Decentralized exchanges (DEXs) are crucial to decentralized finance (DeFi) as they enable trading without intermediaries. However, they face challenges like impermanent loss (IL), where liquidity providers (LPs) see their assets' value change unfavorably within a liquidity pool compared to outside it. To tackle these issues, we propose dynamic fee mechanisms over traditional fixed-fee structures used in automated market makers (AMM). Our solution includes asymmetric fees via block-adaptive, deal-adaptive, and the "ideal but unattainable" oracle-based fee algorithm, utilizing all data available to arbitrageurs to mitigate IL. We developed a simulation-based framework to compare these fee algorithms systematically. This framework replicates trading on a DEX, considering both informed and uninformed users and a psychological relative loss factor. Results show that adaptive algorithms outperform fixed-fee baselines in reducing IL while maintaining trading activity among uninformed users. Additionally, insights from oracle-based performance underscore the potential of dynamic fee strategies to lower IL, boost LP profitability, and enhance overall market efficiency.

\end{abstract}


\section{Introduction}
\label{sec:intro}

Decentralized exchanges (DEXs) are integral to decentralized finance (DeFi), enabling trading without intermediaries~\cite{malamud2017decentralized,zetzsche2020decentralized}. Despite advantages over centralized exchanges (CEXs), such as eliminating middlemen~\cite{barbon2021quality}, DEXs face challenges like Impermanent Loss (IL) and arbitrage~\cite{wang2022cyclic}. IL occurs when liquidity providers (LPs) see a reduction in asset value within a liquidity pool compared to holding them outside, especially in volatile markets, diminishing their incentive to provide liquidity. Transaction fees can partly offset these losses, providing some compensation for LPs. Arbitrage exploits price discrepancies between exchanges, often exacerbating IL and destabilizing DEX market efficiency~\cite{hagele2024centralized}.

To address these issues, dynamic fee mechanisms~\cite{Nezlobin2023} have been proposed as alternatives to fixed fees in automated market makers (AMMs)~\cite{pourpouneh2020automated}. These fees adjust to market conditions such as volatility and trading volume, potentially mitigating excessive arbitrage and IL. By increasing fees during high volatility, DEXs can capture more value from arbitrage trades, compensating LPs for increased risk.

Recent studies indicate a rise in DEX popularity, with platforms like Uniswap, PancakeSwap, and SushiSwap gaining traction~\cite{nartey2024rise,makridis2023rise,hagele2024centralized}. Research highlights risks faced by LPs, particularly IL~\cite{capponi2021adoption,heimbach2021behavior, heimbach2022risks, aigner2021uniswap}. Solutions to IL include Impermanent Gain products~\cite{bardoscia2023liquidity} and frameworks for generalizing IL across exchanges~\cite{tangri2023generalizing}. Arbitrage between DEXs and CEXs remains a significant IL factor, with studies exploring architectural differences and arbitrage strategies~\cite{jansen2023secure,boonpeam2021arbitrage}.

Our contributions are: (i) a mathematical model for optimal fee structures in AMMs based on market participant preferences; (ii) exploration of dynamic fee mechanisms to mitigate IL and reduce arbitrage inefficiencies; (iii) a framework for comparing fee structures, considering LP IL, DEX user profits, and market scenarios; and (iv) an algorithm for dynamic fee setting based on a CEX price oracle, providing insights into potential IL reduction.

The code accompanying this study is publicly available on GitHub~\cite{OurGithub}.

\section{DEX User Exchange Model}
\label{sec:Scenario}
This study aims at solving real-world problems by testing the proposed methods in simulations that are close to real-world scenarios. To this end, we focus on two primary types of users commonly observed in DEXs: liquidity providers, who provide liquidity to earn transaction fee, and traders who exchange crypto assets. For the purpose of this study, traders are further classified into two distinct groups: (i) \textbf{informed users~(IU)}, who execute trades to maximize their profit, and (ii) \textbf{uninformed users~(UU)}, who make trades based on internal principles, which may not always be optimal. Informed users are modeled through profitable arbitrage opportunities with CEXs, while uninformed users conduct more random transactions with a tolerance for inefficiency.

CEX asset prices are considered the ground truth in this study. CEX users operate using an order book, and we define the true price as the mean of the highest bid and lowest ask prices. While there are multiple CEXs, each with its own order book, price deviations are typically smaller on CEXs due to higher liquidity. These characteristics motivate our model, which assumes a single aggregated CEX and one DEX liquidity pool.

The DEX pool is defined by a token pair, which we will call \textbf{Token A} and \textbf{Token B}. At any given time $t$ the pool contains $x$ units of Token A and $y$ units of Token B. The DEX exchange rate for Token B in terms of Token A, $p_{DEX}$, is a function of $x$ and $y$: $p_{DEX} = p_{DEX}(x, y)$. The DEX operates under an automated market maker mechanism, characterized by an invariant function $I(x, y)$, which determines both the pricing and the allowable trade amounts: $(\Delta x, \Delta y) = \left(\Delta x(x, y, \Delta y), \Delta y(x, y, \Delta x)\right),$
where $(\Delta x, \Delta y)$ are the amounts of the exchanged tokens A and B by a user. The values of $(\Delta x, \Delta y)$ are determined by the invariant function and fee policy. The sign of a component determines the exchange direction: positive values correspond to tokens bought from the pool, while negative values correspond to tokens sold to the pool. In this paper, the constant product automated market maker~(CPMM) is considered. Therefore, the price curve for this AMM type adheres to the equation $I(x, y) = x\cdot y$.


Transaction fees in the pool are represented as $(x_f, y_f) \\= f(x, y, \Delta x, \Delta y)$.

The asset prices on CEX in a base currency are denoted as $p_A=p_A(t)$ and $p_B=p_B(t)$ and the exchange rate is given by $p_{CEX}=p_B/p_A$.

Let us consider each type of user's behavior in more detail. The behavior of users is influenced by the current state of the DEX pool and token prices on CEX. The \textbf{capital function} is defined as follows: $P(a, b) = a\cdot p_A + b\cdot p_B$,
where $(a,b)$ represents an arbitrary pair of real numbers.

At the moment $t$ LPs have capital $P(x, y) = x\cdot p_A + y\cdot p_B$. This capital is distributed among LPs via liquidity tokens. Liquidity providers aim to maximize their returns through transaction fees. While fees increase LPs' capital, imbalances in the token quantities $x$ and $y$ can lead to IL. Such imbalances may also reduce overall liquidity in the pool, potentially making the pool less attractive to users.
Users pay a $\alpha = \alpha(t)$ for exchange transaction to the network. The change in an users' capital, $\delta P$, is given by: 
\begin{eqnarray}
    \label{eq:userCapitalChane}
    \delta P = P(\Delta x, \Delta y) - P(f(x, y, \Delta x, \Delta y)) - \alpha.
\end{eqnarray}

\subsection{Informed User}
\label{subsec:IU}
As previously mentioned, the informed user executes a transaction only when it will bring a profit. The IU aims to exchange a specific amount of tokens $\Delta x, \Delta y$ to maximize the change in the capital function $\delta P$ over $(\Delta x, \Delta y)$.

If the maximum of $\delta P$ is positive, the IU broadcasts the exchange transaction with corresponding optimal amounts $(\Delta x^*, \Delta y^*)$. 

Let us illustrate it with an example of the CPMM model with fixed fee $f_{\text{fx}}$.
Suppose IU wants to exchange $\Delta x$ tokens of Token A to $\Delta y$ tokens of Token B. The invariant equation for the pool is expressed as:
\begin{equation}
    \label{eqn:I}
    I(x,y) = x\cdot y = (x + \Delta x \cdot (1-f_{\text{fx}}))\cdot (y - \Delta y).
\end{equation}

The IU’s goal is to maximize the change in their capital function $\delta P$. By rearranging (\ref{eqn:I}) to express $\Delta y$ as a function of $\Delta x$, and substitutiong in (\ref{eq:userCapitalChane}) the IU will receive:
\begin{equation}
    \Delta x^* = \left(\sqrt{x\cdot y\cdot (1-f_{\text{fx}})/p_{CEX}}-x\right)/(1-f_{\text{fx}}).
\end{equation}




Once $\Delta x^*$ is determined, the corresponding value of $\Delta y^*$ can be computed using the CPMM invariant~(\ref{eqn:I}). These optimal values allow the IU to execute the most profitable trade, accounting for both the DEX pool’s transaction fees and the price discrepancy between the DEX and the CEX.
By consistently optimizing their trades in this manner, IUs contribute to rebalancing the pool and ensuring arbitrage opportunities are exploited efficiently. However, excessive arbitrage activity may also exacerbate IL for liquidity providers, highlighting the importance of carefully calibrated fee structures to maintain equilibrium in the system.

\subsection{Uninformed User}
\label{subsec:UU}
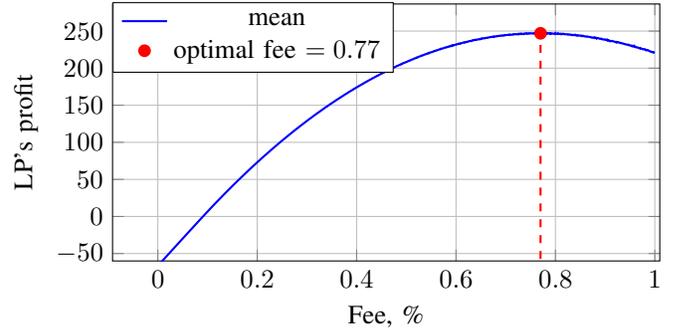
\begin{figure}
    \centering
\begin{tikzpicture}
    \begin{axis}[
        xlabel={Fee, \%},
        ylabel={LP's profit},
        grid=major,
        ymin=-60,
        xmax=1.01,
        ytick={-50, 0, 50, 100, 150, 200, 250},
        width=\columnwidth,
        height=5cm,
        legend style={
            at={(0,0.87)}, 
            anchor=west 
        }    ]
        \addplot[
            thick,
            blue
        ] table[
            x=fee,
            y=profit_avg,
            col sep=comma
        ] {data/f_factor_new1.csv};
        \addlegendentry{mean}

        \addplot[
            only marks,
            mark=*,
            mark options={red},
            thick
        ] coordinates {
            (0.77, 247)
        };
        \addlegendentry{optimal fee $=0.77$}

        \addplot[
            name path=upper,
            draw=none
        ] table[
            x=fee,
            y=std_upper_bound_10,
            col sep=comma
        ] {data/f_factor_new1.csv};
        \addplot[
            name path=lower,
            draw=none
        ] table[
            x=fee,
            y=std_down_bound_10,
            col sep=comma
        ] {data/f_factor_new1.csv};

        \addplot[
            fill=blue!20
        ] fill between[
            of=upper and lower
        ];

        \addplot[
            thick,
            red,
            dashed,
            domain=-60:10
        ] coordinates {
            (0.77, -100) (0.77,255)
        };
    \end{axis}
\end{tikzpicture}
\caption{LP's profit as a function of fee}
\label{fig:r_factor}
\vspace{-0.5cm}
\end{figure}

The primary goal of uninformed users is to exchange one token for another, often with limited sensitivity to small losses incurred during the transaction. However, psychological factors significantly influence the likelihood of a user completing a transaction, especially when the perceived or actual losses become substantial. This insight underscores that setting excessively high transaction fees is not a sustainable solution to address the issue of IL faced by liquidity providers. While high fees may temporarily increase the LP's revenue per transaction, they discourage users from participating in transactions, resulting in lower transaction volume and ultimately lower overall LP profits (see Fig.~\ref{fig:r_factor}).

To capture this behavioral aspect, we introduce a ``psychological'' relative loss factor $r$, which quantifies the ratio of deal losses to transaction volume:
\begin{equation}
\label{eq:r_factor}
    r = \delta P(\Delta x, \Delta y)/\delta P(|\Delta x|, |\Delta y|). 
\end{equation}

This metric reflects the UU's propensity to complete a transaction, depending on the extent of perceived losses relative to the total transaction value. A lower value of $r$ indicates that users perceive the transaction as less favorable, reducing their willingness to proceed.

Uninformed users exhibit a distinct behavior compared to informed users, as they tend to make trades based on internal principles or heuristics, rather than optimizing for arbitrage opportunities. Their actions may follow patterns such as balancing portfolios, rebalancing assets after market shifts, or making trades based on external, non-financial motivations. Unlike informed users, UUs are less likely to evaluate the transaction in terms of maximizing returns and more likely to be influenced by perceived fairness or simplicity of the trade. When $r$ exceeds a tolerable threshold, UUs are less likely to proceed with the exchange.

\section{Fee Algorithms}
\label{sec:FeeAlgorithms} 

To address the issue of IL, we propose the concept of asymmetric or directed fees: fee rates that vary depending on the direction of the swap. Specifically, the fee for swaps from Token A to Token B may be set higher than for swaps in the reverse direction. Ideally, we would set a higher fee in direction of potential arbitrage.

However, this approach is limited by the inherent limitations of DEX in accessing external data. Since fee algorithms in DEX are restricted to using data available on the blockchain, they face a fundamental information asymmetry when compared to arbitrageurs, who can leverage off-chain data sources to identify profitable opportunities. This asymmetry poses a challenge to fee optimization strategies.

In this paper, we evaluate the performance of the following fee algorithms:

\subsection{Fixed (FX, baseline)}

The fixed fee is the traditional approach where the fee rate is constant for both A $\rightarrow$ B and B $\rightarrow$ A swaps, denoted as $f_{\text{fx}}$. For the purposes of our experiments, we set $f_{\text{fx}} = 30 \text{bps}$ (basis points, $0.01\%$).

\subsection{Block-Adaptive~(BA)}

This fee algorithm is based on the observation that if an arbitrage transaction occurs in block $N$, the likelihood of another arbitrage transaction in block $N+1$ in the same direction is higher than in the opposite direction:




\begin{enumerate}
    \item Begin with two separate fee rates for each direction, denoted as $f_{A\rightarrow B}$ and $f_{B\rightarrow A}$.
    
    \item If $q_{\text{DEX}}$ decreases during a block (indicating a net increase in A $\rightarrow$ B transactions and a potential arbitrage opportunity in the A $\rightarrow$ B direction), and $f_{B\rightarrow A} \geq f_{\text{step}}$, increase $f_{A\rightarrow B}$ by $f_{\text{step}}$ and decrease $f_{B\rightarrow A}$ by $f_{\text{step}}$.
    
    \item Similarly, if $q_{\text{CEX}}$ increases during a block (indicating a potential arbitrage opportunity in the B $\rightarrow$ A direction), and $f_{A\rightarrow B} \geq f_{\text{step}}$, increase $f_{B\rightarrow A}$ by $f_{\text{step}}$ and decrease $f_{A\rightarrow B}$ by $f_{\text{step}}$.
    
    \item If the price remains unchanged during the block, fee rates are left unaltered.

\end{enumerate}

\subsection{Deal-Adaptive (DA)}
This algorithm dynamically adjusts the fee based on the direction of the previous transaction. For instance, if the last transaction involved exchanging Token A for Token B, the fee in the direction from A to B ($f_{A\rightarrow B}$) increases by 1 bps, while the fee in the reverse direction ($f_{B\rightarrow A}$) decreases by 1 bps. The algorithm ensures that the average of the fees, $f_{A\rightarrow B}$ and $f_{B\rightarrow A}$, does not exceed 30 bps. Consequently, in cases of frequent transactions in a single direction (an indicator of arbitrage activity), the fee in that direction gradually increases, while it decreases in the opposite direction. This mechanism helps boost the profit of LPs and encourage users to make transactions in the less active direction, improving market balance and efficiency.

\subsection{Oracle-Based (OB)}

The discrete fee algorithm with a perfect oracle represents an idealized scenario designed to establish an upper bound on the effectiveness of fee mechanisms in mitigating IL.

This algorithm assumes access to the exact "fair prices" used by arbitrageurs to calculate their optimal trades, thereby eliminating the information asymmetry. While this assumption is highly unrealistic in practical settings, it serves as a benchmark for comparing more feasible algorithms.

To reduce the attractiveness of arbitrage opportunities, the algorithm dynamically adjusts the fee rates based on the direction of arbitrage. Specifically, the fee for swaps in the "arbitrage" direction (either $A \rightarrow B$ or $B \rightarrow A$) is set to $f_{\text{ad}}$, while the fee in the opposite direction is set to $f_{\text{nad}}$. In our experiments, we use $f_{\text{ad}} = 45 \text{bps}$ and $f_{\text{nad}} = 15 \text{bps}$.

\section{Evaluation Framework for Fee Algorithms}
\label{sec:ComparisonFrameWork}

To systematically compare and evaluate the proposed fee algorithms, we develop a simulation-based framework that emulates trading activity on a DEX at the granularity of individual blockchain blocks.


The initial pool size is set to $25 \text{M}$ USDT, evenly distributed between the two tokens. For each block in the simulation, we model user behavior and trading dynamics as follows:

\begin{enumerate}
    \item \textbf{Uninformed User Activity}: With a probability of $p_{\text{UU}}$, an uninformed user attempts to trade. If a trade occurs, the user selects the direction and amount randomly. The traded amount is modeled as a fraction of the pool’s assets, following a normal distribution $\mathcal{N}(\mu_{\text{UU}}, \sigma_{\text{UU}}^2)$.
    
    \item \textbf{Informed User Arbitrage}: Informed users execute arbitrage transactions whenever profitable opportunities exist, based on the "fair" prices for the block.
    
    \item \textbf{Metrics Calculations}: For all market participants, metrics  are computed relative to the block’s fair prices.
\end{enumerate}

The parameters governing pool dynamics and user behavior are calibrated to closely match the characteristics of the ETH-SHIB liquidity pool on Uniswap V2.

Throughout the simulation, we calculate the markouts (MO), representing the profitability of trades in USDT for each type of participant: IU, UU, and LP. For LPs, the MO is the negative of the IL, adjusted using a hold strategy. A higher markout value indicates better performance, while the smaller IL the better.



The simulation utilizes two types of price data to ensure robustness under diverse market conditions.
\textbf{Synthetic Data}: Simulating near-realistic token prices using geometric Brownian motion. A total of 1,000 paths, each spanning one day (1,440 blocks), were generated. We use 4 sets of parameters for GBM (estimated on corresponding segments of real data)
\textbf{Historical Data}:
Real-world price data for ETH and SHIB is sourced from Binance, with 1-minute snapshots. To account for different market conditions, we simulate four distinct periods: \textbf{Bear Market}~(April 2024), \textbf{Bull Market}~(November 2024), \textbf{High Volatility}~(March 2024), \textbf{Low Volatility}~(August 2024).   
Each period consists of approximately $40,000$ samples, which are treated as individual blocks in the simulation.

\section{Numerical Results}
\label{sec:Results}

\subsection{Psychological Factor Modeling}
\label{subsubsec:r_factor}
As highlighted in Section~\ref{subsec:UU}, excessively high fees do not benefit LPs as they result in reduced transaction volumes, ultimately leading to lower overall profits for LPs. To account for this, our model includes a “psychological factor” $r$ (\ref{eq:r_factor}), which determines the likelihood of a user completing a transaction. To illustrate this effect, we conducted simulations using the following setup: we fixed the quantities of tokens A and B in the pool ($x, y$), set the number of A tokens to be exchanged ($\Delta x$), and maintained constant CEX prices for tokens A and B. The number of tokens B returned ($\Delta y$) was then calculated for varying fee levels (using (\ref{eqn:I})). Subsequently, the transaction probability was modeled as $P = \exp(-|r|)$. Figure~\ref{fig:r_factor} presents the results averaged over $10^6$ simulations, demonstrating that at higher fees, LP revenue decreases due to diminished transaction demand.

\begin{table}[ht]
    \centering
    \caption{Synthetic data simulation results}
    \label{tab:joint-pd}
    \begin{threeparttable}
    \begin{tabular}{|c|c|c|c|c|}
    \toprule
    Market & Alg. & IU MO & UU MO & LP MO \\
    \midrule
    \multirow{4}{*}{\rotatebox{90}{\shortstack{High\\Volatile}}} 
    & FX & $85640 \pm 8477$ & $\text{-}8768 \pm 611$ & $\text{-}84133 \pm 8222$ \\
    & DA & $85620 \pm 8478$ & $\mathbf{\text{-}8766 \pm 615}$ & $\text{-}84115 \pm 8215$ \\
    & BA & $\mathbf{84742 \pm 8409}$\tnote{*} & $\text{-}8773 \pm 587$ & $\mathbf{\text{-}83220 \pm 8167}$ \\
    & OB & $69766 \pm 7414$ & $\text{-}8804 \pm 563$ & $\text{-}67541 \pm 7193$ \\
    \midrule
    \multirow{4}{*}{\rotatebox{90}{\shortstack{Low\\Volatile}}} 
    & FX & $595 \pm 103$ & $\text{-}8116 \pm 284$ & $3450 \pm 226$ \\
    & DA & $597 \pm 101$ & $\text{-}8115 \pm 284$ & $3447 \pm 223$ \\
    & BA & $\mathbf{591 \pm 99}$ & $\mathbf{\text{-}8113 \pm 279}$ & $\mathbf{3460 \pm 219}$ \\
    & OB & $434 \pm 93$ & $\text{-}8118 \pm 279$ & $3742 \pm 224$ \\
    \midrule
    \multirow{4}{*}{\rotatebox{90}{Bull}} 
    & FX & $2976 \pm 341$ & $\mathbf{\text{-}8150 \pm 317}$ & $502 \pm 407$ \\
    & DA & $\mathbf{2982 \pm 344}$ & $\text{-}8152 \pm 318$ & $499 \pm 409$ \\
    & BA & $2909 \pm 338$ & $\text{-}8155 \pm 308$ & $\mathbf{596 \pm 401}$ \\
    & OB & $2207 \pm 312$ & $\text{-}8160 \pm 291$ & $1557 \pm 389$ \\
    \midrule
    \multirow{4}{*}{\rotatebox{90}{Bear}} 
    & FX & $1193 \pm 170$ & $\text{-}8116 \pm 294$ & $2640 \pm 281$ \\
    & DA & $1192 \pm 169$ & $\text{-}8117 \pm 294$ & $2642 \pm 282$ \\
    & BA & $\mathbf{1167 \pm 166}$ & $\mathbf{\text{-}8112 \pm 286}$ & $\mathbf{2679 \pm 264}$ \\
    & OB & $872 \pm 158$ & $\text{-}8105 \pm 288$ & $3139 \pm 278$ \\
    \bottomrule
    \end{tabular}
    \begin{tablenotes}
    \item[*] For each metric, the best algorithms among \texttt{FX, DA}, and \texttt{BA} are in bold.
    \end{tablenotes}
    \end{threeparttable}
    \vspace{-0.5cm}
\end{table}

\begin{table}[ht]
    \centering
    \caption{Historical data simulation results}
    \label{tab:combined-market}
    \begin{tabular}{|c|c|c|c|c|c|}
    \toprule
    Market & Alg. & IU MO & UU MO & LP MO& IL \\
    \midrule
    \multirow{4}{*}{\rotatebox{90}{\shortstack{High\\Volatile}}} 
    & FX & 883,900 & \textbf{-314,528} & -713,551 & 2,146,636 \\
    & DA & 883,295 & -315,492 & -711,763 & 2,145,809 \\
    & BA & \textbf{878,088} & -319,057 & \textbf{-702,735} & \textbf{2,140,245} \\
    & OB & 745,471 & -318,141 & -562,275 & 1,995,396 \\
    \midrule
    \multirow{4}{*}{\rotatebox{90}{\shortstack{Low\\Volatile}}} 
    & FX & 21,547 & \textbf{-222,801} & 86,924 & -129,736 \\
    & DA & \textbf{21,326} & -224,479 & \textbf{88,922} & \textbf{-130,708} \\
    & BA & 21,556 & -222,936 & 86,584 & -130,465 \\
    & OB & 17,768 & -223,127 & 93,014 & -135,422 \\
    \midrule
    \multirow{4}{*}{\rotatebox{90}{Bull}} 
    & FX & 183,885 & -272,720 & -37,780 & -497,942 \\
    & DA & 183,091 & -273,177 & \textbf{-36,148} & \textbf{-512,479} \\
    & BA & \textbf{180,639} & \textbf{-269,788} & -36,720 & -505,210 \\
    & OB & 148,360 & -271,359 & 2,719 & -541,723 \\
    \midrule
    \multirow{4}{*}{\rotatebox{90}{Bear}} 
    & FX & 63,221 & \textbf{-221,017} & 43,210 & \textbf{-183,518} \\
    & DA & \textbf{61,989} & -222,020 & \textbf{45,000} & -182,912 \\
    & BA & 63,081 & -222,168 & 44,126 & -181,757 \\
    & OB & 53,033 & -222,757 & 58,253 & -192,818 \\
    \bottomrule
    \end{tabular}
    \vspace{-0.5cm}
\end{table}

\subsection{Synthetic Data}
\label{subsec:synthetic_data}

We evaluate fee algorithms using synthetic data simulating various market conditions: high/low volatility, bull, and bear markets (Table \ref{tab:joint-pd}). Our goal is to analyze result variability and establish statistical significance. LP's IL is excluded due to high variance.

The \texttt{OB} algorithm consistently surpasses the \texttt{FX} baseline, demonstrating the benefits of fixed fee strategies. In all conditions, \texttt{BA} outperforms \texttt{FX} in LP Markout, proving its effectiveness, though \texttt{OB} yields superior results. The \texttt{DA} algorithm shows minor improvements over \texttt{FX}, but is less effective compared to \texttt{BA}.

\subsection{Historical Data}
\label{subsec:historical_data}

We evaluate fee algorithms under various market conditions, comparing MOs and LP's IL to the \texttt{FX} baseline (Table~\ref{tab:combined-market}).

In a \textbf{high volatile market}, proposed algorithms reduce IU markout significantly compared to \texttt{FX} (883,900), with \texttt{BA} (878,088) and \texttt{OB} (745,471) showing effectiveness. LP markout improves with adaptive algorithms; \texttt{BA} enhances LP profitability to -702,735 from -713,551. \texttt{OB} minimizes IL from 2,146,636 to 1,995,397 in high-volatility environments.

Under \textbf{low volatile market}, the \texttt{DA} algorithm significantly improves metrics, reducing IU profit from 21,547 to 21,326 and increasing LP profit by 2.3\% (from 86,924 to 88,922). The \texttt{OB} method further decreases IU markouts by 17\% and boosts LP markouts by 7\%. Overall, all algorithms (\texttt{DA, BA, OB}) enhance IL.

In a \textbf{bull market}, changes in metrics are minor. \texttt{OB} achieves a positive LP markout of 2,719, setting a benchmark for LP profitability. \texttt{FX} records the highest IL (-497,942), while \texttt{DA} and \texttt{BA} reduce IL by 3\% and 1.5\%, respectively, improving LP markout slightly.

In a \textbf{bear market}, dynamic fee strategies prove resilient. IU markout decreases across all algorithms, with \texttt{OB} achieving superior LP markout (58,253). \texttt{DA} and \texttt{BA} improve LP markout to 45,000 and 44,126, outperforming \texttt{FX} (43,210) by 4\% and 2\%, respectively.

\section{Conclusions and future work}
\label{section:Conclusions}


This study shows that dynamic fee mechanisms, including block-adaptive, deal-adaptive, and oracle-based algorithms, effectively reduce impermanent loss (IL) for liquidity providers on decentralized exchanges (DEXs), outperforming traditional fixed fee models. These dynamic fees adapt to market volatility, compensating LPs during high volatility and reducing inefficiencies and arbitrage between DEXs and CEXs, thus enhancing market efficiency.

Future research should test these algorithms in live DEX environments to verify theoretical results and uncover practical challenges. Incorporating machine learning could further refine fee adaptability by predicting market conditions and user behavior. Additionally, exploring arbitrage opportunities and LP and trader responses to dynamic fees can inform improved algorithm designs that balance market efficiency with user satisfaction.

\bibliographystyle{IEEEtran}
\bibliography{bibliography}

\end{document}